\documentclass{ifacconf}

\usepackage{graphicx}      
\usepackage{natbib}        
\usepackage{amsmath,amsfonts,amssymb,amscd,latexsym,enumerate,bbm,stfloats}
\usepackage{amssymb}
\usepackage{comment}
\usepackage{color}
\usepackage{subeqnarray}
\usepackage{algorithmic}
\usepackage{enumitem}
\usepackage[short]{optidef}
\usepackage{theorem}

\newcommand{\calS}{\ensuremath{\mathcal{S}}}

\newcommand{\calN}{\ensuremath{\mathcal{N}}}
\newcommand{\calK}{\ensuremath{\mathcal{K}}}
\newcommand{\calZ}{\ensuremath{\mathcal{Z}}}

\newcommand{\calQ}{\ensuremath{\mathcal{Q}}}

\newtheorem{problem}{Problem}
\newtheorem{definition}{Definition}
\newtheorem{remark}{Remark}
\newtheorem{lemma}{Lemma}
{\theorembodyfont{\itshape}\newtheorem{theorem}{Theorem}}

\theoremheaderfont{\normalfont\bfseries}

\begin{document}
	\begin{frontmatter}
		
		\title{Data-driven Linear Quadratic Regulation via Semidefinite Programming}
		
		\thanks[footnoteinfo]{This project is conducted under the auspices of the Centre for Data Science and Systems Complexity at the University of Groningen and is supported by  a Marie Sk\l{}odowska-Curie COFUND grant, no. 754315.}
		
		\author[First]{Monica Rotulo} 
		\author[First]{Claudio De Persis} 
		\author[Second]{Pietro Tesi}
		
		\address[First]{ENTEG, University of Groningen, 9747 AG Groningen, The Netherlands (email: \{m.rotulo, c.de.persis\}@rug.nl).}
		\address[Second]{DINFO, University of Florence, 50139 Florence, Italy
			(email: pietro.tesi@unifi.it)}
		
		\begin{abstract}          
			This paper studies the finite-horizon linear quadratic regulation problem 
			where the dynamics of the system are assumed to be unknown and the state is accessible. 
			Information on the system is given by a finite set of input-state data, where the input 
			injected in the system is persistently exciting of a sufficiently high order. 
			Using data, the optimal control law is then obtained as 
			the solution of a suitable semidefinite program. 
			The effectiveness of the approach is illustrated via numerical examples.
		\end{abstract}
		
		
	\end{frontmatter}
	
	\section{Introduction}
	Optimization and control have always been closely
	 related when there is need for operating a dynamical system at minimum cost.  
	When the system is linear and the cost function is quadratic, the optimal control 
	problem amounts to solving the popular linear quadratic regulator (LQR) problem \citep{anderson2007optimal}, 
	whose duality with convex optimization is shown in \cite{balakrishnan1995connections} 
	for continuous-time systems and more recently in \cite{gattami, lee} for discrete-time systems. 
	The common assumption for deriving the solution is that 
	an exact model of the system is available.
	
	To cope with the lack of prior knowledge of the system dynamics, 
	various control techniques have been developed. The classic approach is 
	the so-called \emph{indirect} approach:
a model is first determined from data, and then the control law is designed using the model. 
It is worth to mention that the control objectives are not taken into account in the identification step and, 
once the model is derived from data, the data is not used in the synthesis of the control law. 
More recently, as opposed to the 
model-based paradigm, \emph{data-driven} control, also termed 
\emph{model-free}, has become increasingly popular. 
Data-driven control is based on the paradigm of learning a controller directly 
from data
collected from the system.
Various efforts have been made in this direction, in the context of optimal control  for linear systems
\citep{Markovsky_LQR,da2018data, baggio2019data, coulson2019data} as well as for designing control laws for nonlinear systems
\citep{fliess,koopman,safonov}.
Data-driven control has also been approached using popular Machine Learning  
tools such as Reinforcement Learning (RL).
In RL \citep{RL_sutton, hagen,  barto}, 
also referred to as approximate dynamic programming \citep{Lewis_RL,busoniu2017reinforcement}, 
the controller learns an optimal policy through trial-and-error, 
trying to estimate a long-term value function. Other data-driven techniques are discussed in the surveys  \cite{benosman2018model,hou_survey} to which the interested reader is referred.
Despite advances in this area,
data-driven control still poses many theoretical and practical challenges. 
For instance, 
in applications involving on-line control design, approaches like
RL might suffer from the large number of iterations required to achieve convergence 
\citep{gorges2017relations}. 
	
	In this paper, we consider the \emph{finite-horizon} LQR problem 
	for linear time-invariant discrete-time systems. Different from the mainstream 
	approaches based on RL techniques \citep{zhao2015finite, pang, liu_basar},
	the approach we propose does not involve iterations. We 
	formulate the LQR problem as a (one-shot) semidefinite program 
	in which the model of the system is replaced by a finite number of data
	collected from the system.
	This idea has been proposed in \cite{cp} for the infinite-horizon LQR problem.
	The method proposed in this paper also recovers the infinite-horizon solution in a very natural way.
	
	Our approach is based on the framework developed in \cite{cp}, 
	whose foundation lies on the \textit{fundamental lemma} by \cite{willems2005note}. 
	Roughly speaking, the \textit{fundamental lemma} stipulates that 
	one can describe all possible trajectories of a linear time-invariant system using any given finite set of its
	input-output data, 
	provided that these data come from sufficiently excited dynamics. 
	This result thus establishes that data implicitly give a \emph{non-parametric}
	system representation which can be directly used for control design.
	
	The paper is organized as follows: Section \ref{sec:modelbased} 
	briefly reviews the model-based finite-horizon LQR 
	solution. In Section 3, the LQR problem is reformulated as a convex optimization 
	problem involving linear matrix inequalities \citep{boyd1994linear}, resulting in a semidefinite program (SDP). 
	The proposed approach is discussed in Section \ref{sec:DD}. We show that 
	the data-based parametrization introduced in
	\cite{cp} combined with the SDP formulation of the LQR problem 
	results in a direct parametrization of the feedback system through data. In turn, this
	makes it possible to determine the optimal control law in one-shot,
	with no intermediate identification step. 
	Numerical examples are discussed in Section 4. 
	The paper ends with some concluding remarks  in Section \ref{sec:concl}. 
	
	\textit{Notation:} Given a signal $z: \mathbb{Z} \rightarrow \mathbb{R}^\sigma$, 
	we will denote by $z_{[i,k]}$ the sequence $\{z(i), \dots, z(k) \}$, 
	where $i\leq k$. For a square matrix $A$, we denote by $\mathbf{Tr}(A)$ its trace. 
	In addition, we write $A\succ 0 $ and $ A\succeq 0 $  
	to denote that $A$ is positive definite (semi-definite).
	We use the notation $w(k) \sim \calN(0,W)$ to represent a zero-mean Gaussian random vector such that 
	$\mathbf{E}[w(k)]=0$ and $\mathbf{E}[w(k)w^\top(k)]=W$,
	where $\mathbf{E}$ denotes the expectation.

	\section{Finite-horizon LQR problem}\label{sec:modelbased}
	Consider a discrete-time linear system 
	\begin{equation}\label{sys}
	x(k+1)= Ax(k)+Bu(k)
	\end{equation}
	
	where $x \in \mathbb{R}^n$ is the state while $u \in \mathbb{R}^m$ is the control input, and where
	$A$ and $B$ are matrices of an appropriate dimension.
	It is assumed throughout the paper that $(A,B)$ is 
	controllable and the state is available for measurements.
	Given an initial condition $x(0)=x_0$ and a control sequence 
	$u_{[0,N-1]} =\{u(0),\ \dots, \ u(N-1)\}$
	over the horizon $N \in \mathbb{N}$, we consider the quadratic cost $J$
	associated to system \eqref{sys} starting at $x_0$,
	\begin{equation}\label{prob_s1}  
	J :=   x(N)^\top Q_fx(N) +\sum_{k=0}^{N-1} \rho(u(k),x(k))
	\end{equation}
	where 	
	\[
	\rho(u,x) = x^\top Q_x x+u^\top R u,
	\]
	
	where
	 $Q_x,Q_f\succeq0$ and $R\succ 0$.
	 The finite-horizon linear quadratic regulator (LQR) problem is as follows:
	 
	\begin{problem}\label{lqr_det}
		Given system \eqref{sys} with initial condition $x_0$, and given a time horizon 
		of length $N$, find an input 		sequence 
		 such that the cost function \eqref{prob_s1} is minimized, i.e. solve the minimization problem:
		\begin{eqnarray}\label{min_cost_s}
		\begin{array}{rl}
		\min_{\mu_k} & \, \, J  \nonumber \\[0.3cm]
		\text{subject to  } & \,\, x(k+1) = A x(k) + B u(k) \nonumber \\[0.1cm]
		& \,\, u(k) = \mu_k(x(0),x(1),\ldots,x(k)).
		\end{array}
		\end{eqnarray}
		
		Here, the last constraint means that the control input is a causal function of the system state.
	\end{problem}
	
	The following result holds. 
	
	\begin{lemma}
		For Problem \ref{lqr_det}, the optimal control sequence 
		$u^{*}_{[0,N-1]}= \{u^{*}(0), \ \dots, \ u^{*}(N-1)\}$ is unique, and it
		is generated by the feedback law
		\begin{equation}\label{opt_u0}
		u^{*}(k)=K^{*}(k)x(k)
		\end{equation}
		
		where 
		\begin{equation}\label{K_lqr}
		K^{*}(k) := -(R+B^\top P(k+1)B)^{-1}B^\top P(k+1)A
		\end{equation}
		
		where $P(k)$ is the solution to the so-called discrete-time difference Riccati equation
		\begin{eqnarray*}\label{riccati_eq0}
		\begin{split}
		P&(k) = Q_x+ A^\top P(k+1)A -\\
		& A^\top P(k+1)B(R+B^\top P(k+1)B)^{-1}B^\top P(k+1)A 
		\end{split}
		\end{eqnarray*}
		
		initialized from $P(N) = Q_f$.
	\end{lemma}

	\begin{pf} See for instance \cite{bertsekas2007dynamic}. \hfill $\Box$
			\end{pf}
The computed control law \eqref{K_lqr} is time-varying and defined in the interval $[0,N]$. However, the computation of the gain $K^*(k)$ does not require the knowledge of the current state, and can be computed offline.
	For $N\rightarrow \infty$, if the pair $(Q_x,A)$ is observable,
	the sequence of the matrices 
	$P(k)$ converges to a matrix $P$, which is the so-called \emph{stabilizing}
	solution of the discrete-time algebraic Riccati equation 
	\begin{equation*}
		P = Q_x + A^\top P A  -A^\top PB(R+B^\top PB)^{-1}B^\top PA.
	\end{equation*}
	
	In this case, the optimal control for the infinite-horizon problem is a time-invariant state-feedback 
	$u(k)=K^*x(k)$ with $K^*= -(R+B^\top PB)^{-1}B^\top PA$. 

\begin{remark}
	Here, similarly to the data-driven infinite-horizon LQR problem studied in \cite{cp}, we have assumed that the pair $(A,B)$ is controllable.
	As discussed in Section 3, this assumption ensures that 
	we can always collect sufficiently rich data by applying 
	exciting input signals. 
	As shown in \cite{henk}, except for pathological cases, 
	data richness is indeed necessary for 
	the data-driven solution of the LQR problem. 
	On the other hand,  
	data richness is also necessary for
	reconstructing the system matrices $A$ and $B$
	from data, thus necessary also for the model-based solution whenever 
	$A$ and $B$ have to be identified from data.
\end{remark}

\subsection{Solution as a covariance optimization problem}

For reasons which will become clear in the next section,  
we introduce an equivalent formulation of the LQR problem, where by ``equivalent"
we mean that the corresponding optimal solution is still given by \eqref{opt_u0}.

Consider the linear quadratic stochastic control problem \citep{gattami}:
		\begin{eqnarray}\label{min_cost_s_gattami}
		\begin{array}{rl}
		\min_{\mu_k} & \, \, \mathbf{E}[J] \nonumber \\[0.3cm]
		\text{subject to  } & \,\, x(k+1) = A x(k) + B u(k) + w(k) \nonumber \\[0.1cm]
		& \,\, x(0) \sim \calN(0,I_n)  \nonumber \\[0.1cm]
		& \,\, w(k)\sim \calN(0,I_n)  \nonumber \\[0.1cm]
		& \,\, \mathbf{E}[w(k)x^\top(l)] = 0, \quad \forall l \leq k \nonumber \\[0.1cm]
		& \,\, u(k) = \mu_k(x(0),x(1),\ldots,x(k))
		\end{array}
		\end{eqnarray}
with $J$ as in \eqref{prob_s1}.
As detailed in \cite{gattami}, this problem 
is equivalent to the covariance selection problem
\begin{eqnarray}\label{min_cost_s_gattami_covariance}
		\begin{array}{rl}
		& \min_{V(0), \ldots,V(N) \succeq 0}   \ \mathbf{Tr}\,(Q_f S(N))
		+ \sum_{k=0}^{N-1} \mathbf{Tr} (  Q_x S(k) + R \, U(k)  ) \\[0.3cm]
		& \qquad \text{subject to  }  \\[0.2cm]
		& \qquad S(0) = I_n \\[0.1cm] 
		& \qquad S(k+1) - \begin{bmatrix} A & & B \end{bmatrix} V(k) 
		\begin{bmatrix} A & & B \end{bmatrix}^\top - I_n = 0  \\[0.1cm]
		\end{array} 
		\end{eqnarray}

		for $k=0,\ldots,N-1$, where 
\begin{eqnarray}\label{}
V(k) = \begin{bmatrix} S(k) & & Y(k) \\ Y(k)^\top & & U(k) \end{bmatrix}
:= \mathbf{E} \begin{bmatrix} x(k) \\ u(k) \end{bmatrix}
\begin{bmatrix} x(k) \\ u(k) \end{bmatrix}^\top.
\end{eqnarray}

In particular, the following result holds.  
\begin{theorem} \cite[Theorem 2]{gattami} \label{thm:gattami}
The optimal solution of the covariance selection problem \eqref{min_cost_s_gattami_covariance} 
is given by 
\begin{eqnarray}\label{}
&& S(0) = I_n \nonumber \\[0.1cm]
&& Y(k)^\top = L(k) S(k) \nonumber \\[0.1cm]
&& U(k) = Y(k)^\top S^{-1}(k) Y(k) \nonumber \\[0.1cm]
&& S(k+1) = \begin{bmatrix} A & & B \end{bmatrix} V(k) 
\begin{bmatrix} A & & B \end{bmatrix}^\top + I_n .\nonumber
\end{eqnarray}
		
The corresponding optimal control law is 
$u(k) = L(k) x(k)$ with $L(k)=K^*(k)$ as in \eqref{opt_u0}.
\end{theorem}

\section{Data-driven LQR via Semidefinite Programming}

Building on the formulation described in Section 2.1, it is possible
to derive a simple solution to the LQR problem where the system
matrices $A$ and $B$ are replaced by data. We first show
how the covariance optimization problem can be restated 
in terms of a semidefinite program. Then, we consider a 
parametrization of the feedback system which
results in a pure data-driven formulation.

\subsection{Semidefinite Programming formulation}

The following result holds.

\begin{theorem}\label{eq2ineq}
		The optimal control law for problem \eqref{min_cost_s_gattami_covariance} can be
		computed as the solution $\calK$ to the problem 
\begin{eqnarray}\label{sdp-alt1_ineq} 
		\begin{array}{rl}
		& \min_{\calS, \calK, \calZ}   \ \mathbf{Tr} ( Q_f S(N) )
		+ \sum_{k=0}^{N-1} \mathbf{Tr} (  Q_x S(k) + Z(k)  ) \\[0.3cm]
		& \,\,\,\, \text{subject to  }  \\[0.3cm]
		& \,\,\,\, S(0) \succeq I_n \\[0.1cm] 
		& \,\,\,\, S(k+1) - (A+BK(k)) S(k) (A+BK(k))^\top - I_n \succeq 0 \\[0.1cm]
		& \,\,\,\, Z(k) - R^{1/2}K(k) S(k) K(k)^\top R^{1/2} \succeq 0 \\[0.1cm]
		\end{array}
		\end{eqnarray}

		for $k=0,\ldots,N-1$, where
		\begin{align*}
		\calS &:= \{S(1),\dots, S(N)\} \\[0.1cm]
		\calK &:= \{K(0),\dots, K(N-1)\} \\[0.1cm]
		\calZ &:= \{Z(0),\dots, Z(N-1)\}
		\end{align*}
	\end{theorem} 
	
	\begin{pf}
	Exploiting the fact that the optimal control law takes the form $u(k)=K(k)x(k)$,
	the term $\mathbf{Tr} ( R\, U(k) )$ appearing in the objective function of \eqref{min_cost_s_gattami_covariance} 
	can be written as
	\begin{eqnarray*}
	\mathbf{Tr} ( R \, U(k) )&=& \mathbf{Tr} \, ( R \, \mathbf{E} [K(k) x(k)  x(k)^\top K(k)^\top] )\nonumber \\
	&=& \mathbf{Tr} \, ( R^{1/2} K(k) S(k) K(k)^\top R^{1/2} ).
	\end{eqnarray*}
	
	In addition, 
\begin{eqnarray*}\label{}
V(k) = \begin{bmatrix} I \\ K(k) \end{bmatrix} S(k)
\begin{bmatrix} I \\ K(k) \end{bmatrix}^\top
		\end{eqnarray*}
so that the second constraint of \eqref{min_cost_s_gattami_covariance} 
becomes
	\[
	S(k+1) - (A+BK(k)) S(k) (A+BK(k))^\top + I_n = 0.
	\]
	
	Accordingly, the optimization problem \eqref{min_cost_s_gattami_covariance} 
	is equivalent to the following problem:
	\begin{eqnarray}\label{sdp-alt1_eq}  
		\begin{array}{rl}
		& \min_{\calS, \calK, \calZ}  \ \mathbf{Tr} (Q_f S(N)) 
		+ \sum_{k=0}^{N-1} \mathbf{Tr} (  Q_x S(k) + Z(k)  ) \\[0.3cm]
		& \,\,\,\,  \text{subject to  }  \\[0.3cm]
		& \,\,\,\,  S(0) = I_n \\[0.1cm] 
		& \,\,\,\, S(k+1) - (A+BK(k)) S(k) (A+BK(k))^\top - I_n  = 0 \\[0.1cm]
		& \,\,\,\,  Z(k) - R^{1/2}K(k) S(k) K(k)^\top R^{1/2} = 0 \\[0.1cm]
		\end{array}
		\end{eqnarray}
	
	Finally, let $(\overline \calS, \overline \calK, \overline \calZ)$ be
	an optimal solution to problem \eqref{sdp-alt1_ineq} and let 
	$(\calS^*, \calK^*, \calZ^*)$ be
	an optimal solution to \eqref{sdp-alt1_eq}, where 
	$\calK^* = \{K^*(0),\dots, K^*(N-1)\}$ is the optimal sequence 
	of state-feedback matrices given by \eqref{opt_u0}.
	Also, denote by $\overline J$ and $J^*$ the corresponding costs.
	Clearly $\overline J \leq J^*$
	since \eqref{sdp-alt1_ineq} has a larger feasible set than \eqref{sdp-alt1_eq}.
	To prove the converse inequality, first note that 
	$\overline{S}(k) \succeq S^*(k)$ and $\overline{Z}(k) \succeq Z^*(k)$
	for all $k \geq 0$. This follows because $\overline{S}(0) \succeq S^*(0)$ and since
	$\overline{S}(k) \succeq S^*(k)$ implies
	\begin{eqnarray*}
	&& (A+BK(k)) \left( \overline{S}(k) - S^*(k)  \right) (A+BK(k))^\top \succeq 0 \\[0.1cm]
	&& R^{1/2}K(k) \left( \overline{S}(k) - S^*(k) \right) K(k)^\top R^{1/2} \succeq 0.
	\end{eqnarray*}
	Substituting $(\overline \calS, \overline \calZ)$ in \eqref{sdp-alt1_ineq} 
	and $(\calS^*, \calZ^*)$ in \eqref{sdp-alt1_eq},
	we thus have $\overline J \geq J^*$ so that $\overline J = J_*$.
	In turn, this implies $\overline \calK = \calK^*$ since the optimal control law
	achieving $J^*$ is unique. 
	\hfill $\Box$
	\end{pf}
	
	Defining $H(k) = K(k) S(k)$ and using the property that $S(k) \succeq I_n$ for every $k \geq 0$,
	problem \eqref{sdp-alt1_ineq} can be 
	converted into a semidefinite program. The idea of resorting to SDP formulations
	has been originally proposed in \cite{feron1992numerical} in the context of 
	model-based LQR, and considered in \cite{cp} in the context of 
	data-driven infinite-horizon LQR.

	\subsection{Data-driven parametrization of LQR}\label{sec:DD}
	
	The formulation \eqref{sdp-alt1_ineq} is very 
	appealing from the perspective of computing the control law using data only
	since the decision variables $\calS$, $\calK$ and $\calZ$ appearing in \eqref{sdp-alt1_ineq}
	enter the problem in a form which permits to write the constraints 
	as data-dependent linear matrix inequalities.
	
	Our approach uses the concept of persistence of excitation, which is recalled
	via the following definitions.
	
	\begin{definition}\label{def:Hankel}
		Given a signal $z \in \mathbb{R}^\sigma$, we denote its Hankel 
		matrix as
		\begin{equation*}
		Z_{i,\ell,j} := \begin{bmatrix}
		z(i) & z(i+1) & \cdots & z(i+j-1)\\
		z(i+1) & z(i+2) & \cdots & z(i+j)\\
		\vdots & \vdots & \ddots & \vdots\\
		z(i+\ell-1) & z(i+\ell) & \cdots & z(i+\ell+j-2)\\
		\end{bmatrix}
		\end{equation*}
		where $i \in \mathbb Z$ and $\ell, j \in \mathbb N$. 
		If $\ell = 1$, 	we denote its Hankel matrix as
		\[
		Z_{i,j} = \begin{bmatrix}
		z(i) & z(i+1) & \cdots & z(i+j-1)
		\end{bmatrix}.
		\]
	\end{definition}
	
	\begin{definition}\label{def:PE}
	The signal $z_{[0,T-1]}: [0,T-1]\cap \mathbb{Z} \to \mathbb R^\sigma$ 
	is said to be persistently exciting of order $\ell$ if the matrix
	$Z_{0,\ell,j}$, with $ j=T-\ell+1 $ has full rank $\sigma \ell$. 
	\end{definition}
	
	For a signal to be persistently exciting of order $\ell$, 
	it must be sufficiently long in the sense that $T \geq (\sigma+1)\ell-1$.  
	
	Consider system \eqref{sys},
	\begin{equation*}
	x(k+1)= Ax(k)+Bu(k)
	\end{equation*} 	
	where $x \in \mathbb R^n$ and $u \in \mathbb R^m$.
	Suppose that we carried out an experiment of duration $T \in \mathbb N$, collecting 
	input and state data $u_{d,[0,T]}$ and $x_{d,[0,T]}$ where the subscript ``$d$"
	denotes \emph{data}. Let the corresponding Hankel matrices be	
	\begin{equation}\label{input-state_partition}
	\begin{split}
	U_{0,T} & := \begin{bmatrix}
	u_d(0) & u_d(1) & \dots & u_d(T-1)
	\end{bmatrix}\\[0.1cm]
	X_{0,T} & := \begin{bmatrix}
	x_d(0) & x_d(1) & \dots & x_d(T-1)
	\end{bmatrix}\\[0.1cm]
	X_{1,T} & := \begin{bmatrix}
	x_d(1) & x_d(2) & \dots & x_d(T)
	\end{bmatrix}.\\[0.1cm]
	\end{split}
	\end{equation}

	\begin{lemma}
		\cite[Corollary 1]{willems2005note} Suppose that system \eqref{sys} is controllable. 
		If the input signal $u_{d,[0,T-1]}$ is persistently exciting of order $n+1$ then
		\begin{equation}\label{rank_cond}
		\begin{split}
		\text{rank} \begin{bmatrix}
		U_{0,T}\\[0.1cm]
		X_{0,T}
		\end{bmatrix} = n+m.
		\end{split} 
		\end{equation}
	\end{lemma}
	
	\begin{remark}
	Condition \eqref{rank_cond} expresses the property that the data content is
	sufficiently rich, and this enables the data-driven solution of the LQR problem (\emph{cf.} Remark 2).
	For a discussion on the types of persistently exciting 
	signals the interested reader is referred to \cite[Section 10]{verhaegen2007filtering}.
	\end{remark}
	
	A straightforward implication of the above result is that any input-state  sequence 
	of the system can be expressed as a linear combination of the collected input-state data.
	As shown in  \cite{cp}, one can use  condition \eqref{rank_cond} also for 
	parametrizing an arbitrary feedback interconnection. To see this, consider 
	an arbitrary matrix $K(k)$, possibly time-varying, of dimension $m \times n$.  
	By the Rouch\'e-Capelli theorem, there exists a $T \times n$ matrix $G(k)$
	solution to
	\begin{align}\label{K_data1}
	\begin{bmatrix}
	K(k)\\[0.1cm] I_n
	\end{bmatrix}=\begin{bmatrix}
	U_{0,T}\\[0.1cm] X_{0,T}
	\end{bmatrix}G(k).
	\end{align}
	
	Accordingly the closed-loop system formed by system \eqref{sys} 
	with $u(k) = K(k) x(k)$ is such that
	\begin{eqnarray} \label{eq_cl_param}
	A+BK(k) &=& \begin{bmatrix}
	B & & A
	\end{bmatrix} 
	\begin{bmatrix}
	K(k)\\[0.1cm] I_n
	\end{bmatrix} \nonumber \\
	&=& \begin{bmatrix}
	B & & A
	\end{bmatrix} 
	\begin{bmatrix}
	U_{0,T}\\[0.1cm] X_{0,T}
	\end{bmatrix}G(k) \nonumber \\
	&=& X_{1,T} G(k)
	\end{eqnarray}
	
	where we used the identity $X_{1,T} = A X_{0,T} + B U_{0,T}$.
	
	Using this result, one can provide a data-based formulation
	of problem \eqref{sdp-alt1_ineq}.
	  
	\begin{theorem}\label{probl_DD_CL}
	Consider system \eqref{sys} along with an experiment of length $T \in \mathbb N$
	resulting in input and state data $u_{d,[0,T]}$ and $x_{d,[0,T]}$, respectively.  
	Let the matrices $U_{0,T}$, $X_{0,T}$ and $X_{1,T}$ be as in \eqref{input-state_partition},
	and suppose that the rank condition \eqref{rank_cond} holds.
	Then, the optimal solution to problem \eqref{sdp-alt1_ineq}, hence to
	Problem 1, is given by
		
		\[
		\calK := \{K(0),\dots, K(N-1)\}
		\]

                 with 
                 \[
                 K(k) = U_{0,T}Q(k)S^{-1}(k)
                 \] 
                 
                 where the matrices $Q(k)$ and $S(k)$ solve the 
                 optimization problem
		\begin{eqnarray}\label{sdp-alt1_ineq_data}  
		\begin{array}{rl}
		& \min_{\calS, \calQ, \calZ}   \ \mathbf{Tr} ( Q_f S(N) )
		+ \sum_{k=0}^{N-1} \mathbf{Tr} (  Q_x S(k) + Z(k)  ) \\[0.3cm]
		& \quad \text{subject to  }  \\[0.3cm]
		& \quad S(0) \succeq I_n \\[0.2cm] 
		& \quad S(k) = X_{0,T}Q(k)  \\[0.2cm] 
		& \quad \begin{bmatrix}
					S(k+1) - I_n & & X_{1,T}Q(k)\\[0.1cm]
					Q^\top(k)X_{1,T}^\top & & S(k)
			\end{bmatrix} \succeq 0 \\[0.5cm]
		& \quad \begin{bmatrix}
					Z(k) & & R^{1/2}U_{0,T}Q(k)\\[0.1cm]
					Q(k)^\top U_{0,T}^\top R^{1/2}  & & S(k)
			\end{bmatrix} \succeq 0 \\[0.5cm]
		\end{array}
		\end{eqnarray}
		for $k=0,\ldots,N-1$, where
				
		\begin{equation}\label{calSQZ}
		\begin{split}
		\calS &:= \{S(1),\dots, S(N)\} \\[0.1cm]
		\calQ &:= \{Q(0),\dots, Q(N-1)\} \\[0.1cm]
		\calZ &:= \{Z(0),\dots, Z(N-1)\}
		\end{split}
		\end{equation}
	\end{theorem}
	
	\begin{pf} 
	We show that the constraints of \eqref{sdp-alt1_ineq}
	can be written as in \eqref{sdp-alt1_ineq_data}. To this end, first note that
	the parametrization \eqref{eq_cl_param} implies that the second constraint of \eqref{sdp-alt1_ineq}
	can also be  written as 
		\begin{equation*}\label{cond1}
		S(k+1) - X_{1,T}G(k) S(k)G(k)^\top X^\top_{1,T} - I_n \succeq 0
		\end{equation*}
		
		where $G(k)$ satisfies \eqref{K_data1}.
		Let now 
		\begin{equation}\label{Q_GS}
		Q(k) := G(k)S(k).
		\end{equation}
		
		Exploiting the fact that $S(k) \succeq I_n$ for every $k \geq 0$
		the second constraint of \eqref{sdp-alt1_ineq}
	becomes

	\begin{equation*}\label{cond1a}
		S(k+1) - X_{1,T}Q(k) S^{-1}(k)Q(k)^\top X^\top_{1,T} - I_n \succeq 0
		\end{equation*}
		
		which is equivalent to the third constraint in \eqref{sdp-alt1_ineq_data}. 
		Along the same lines,
		the third constraint in \eqref{sdp-alt1_ineq} can be written as the 
		fourth constraint in \eqref{sdp-alt1_ineq_data}.  Finally, the optimal solution $\calK := \{K(0),\dots, K(N-1)\}$, with $K(k) = U_{0,T}Q(k)S^{-1}(k)$, is obtained from the first one of \eqref{K_data1} and  \eqref{Q_GS}.
	\hfill $\Box$
	\end{pf}
	
	\begin{remark}
		As $N\rightarrow \infty$ the solution converges to the infinite-horizon steady-state
		solution, which is stabilizing. The infinite-horizon formulation is discussed in \cite{cp}.
	\end{remark}

	\section{Numerical examples}
	
	We consider both the semidefinite programs described in Theorem~\ref{eq2ineq} (model-based) 
	and Theorem~\ref{probl_DD_CL} (data-driven) and compare their performance for randomly generated
	systems and for the batch reactor system.
	
	\subsection{Monte Carlo simulations on random systems}

	We perform Monte Carlo simulations with $N_{trials}=1000$ random
	systems with $3$ states and $1$ input. Simulations are performed 
	in MATLAB.  For each trial, the entries of the 
	system matrices are generated using the command \texttt{randn}
	(normally distributed random number). For each trial, the data 
	are generated by applying a random input sequence of length $T=15$ and 
	random initial conditions, again using the command \texttt{randn}. 
	Using CVX (\cite{grant2008cvx}), we solve the model-based program 
	\eqref{sdp-alt1_ineq} and the data-driven program \eqref{sdp-alt1_ineq_data} 
	for $N=10$ steps with $Q_x=Q_f= I_3$ and  $R= 1$, and measure the resulting optimal costs. 
	
\begin{figure}
	\begin{center}
	\includegraphics[width=8.7cm]{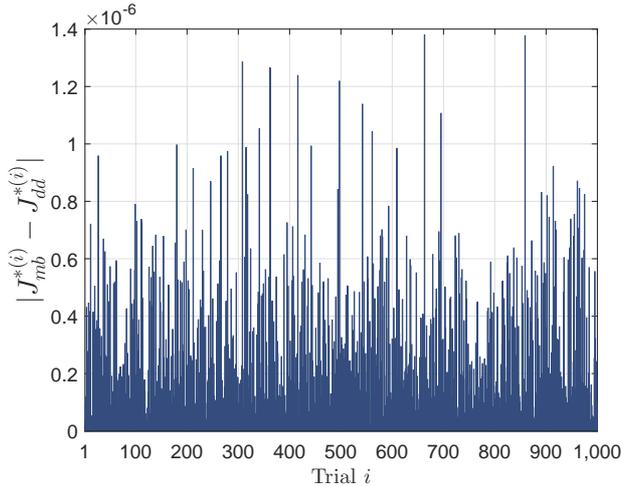}    
	\caption{Absolute error between optimal cost $J^{* (i)}_{mb}$ achieved in model-based and optimal cost $J^{* (i)}_{dd}$ achieved in data-driven for the $i$-th trial.} 
	\label{fig:error_cost1000}
	\end{center}
	\end{figure}

\begin{figure}
	\begin{center}
	\includegraphics[width=8.7cm]{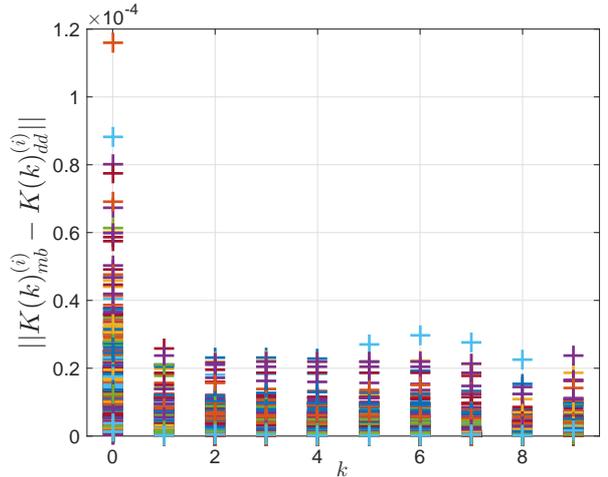}    
	\caption{Values of the error (Euclidean norm) between the optimal model-based solutions $ K(k)^{(i)}_{mb} $ and data-based solutions $ K(k)^{(i)}_{dd} $ for the $i$-th trial, with $ k=0,\dots,~N-1 $. }  
	\label{fig:error_gain1000}
	\end{center}
	\end{figure}

	A shown in Figure~\ref{fig:error_cost1000}, for each trial the data-driven solution achieves
	the same cost as the model-based solution, with an average error of order $10^{-7}$.
	Also the sequence $\calK^*_{dd}$ of data-driven feedback gains coincides with the sequence 
	$\calK^*_{mb}$ obtained by solving the model-based formulation,
	with an average error over the various gains of order $10^{-6}$ 
	(Figure~\ref{fig:error_gain1000}).
	
\subsection{Monte Carlo simulations on batch reactor system}

As a second example, we consider the discretized 
version of the batch reactor system \citep{Walsh2001}, using a sampling time of $0.1s$,
\begin{align} 
& \left[
\begin{array}{c|c}
A & B
\end{array}
\right] = \nonumber \\
& \quad
\left[
\begin{array}{rrrrc|crr}
    1.178  &  0.001 &   0.511 &  -0.403 & & & 0.004 &   -0.087 \\
   -0.051 &    0.661 &   -0.011 &   0.061 & &  &    0.467 &   0.001 \\
    0.076 &    0.335 &    0.560 &    0.382 & & &  0.213 &   -0.235 \\
   0 &    0.335 &   0.089 &   0.849 &  & & 0.213 &   -0.016
\end{array}
\right] \nonumber
\end{align}
	which is open-loop unstable. 
	
	Under the same experimental conditions as in the previous 
	example ($T=15$, $N=10$, $N_{trials}=1000$), and taking
	cost weights $Q_x=Q_f= I_4$ and  $R= I_2$, Monte Carlo simulations return 
	an average error of order $10^{-3}$
	for what concerns the discrepancy $|J^*_{dd}-J^*_{mb}|$, and
	an average error of order $10^{-3}$
	for what concerns $\|\calK^*_{dd}-\calK^*_{mb}\|$.
	
\begin{figure}
	\begin{center}
		\includegraphics[width=8.8cm]{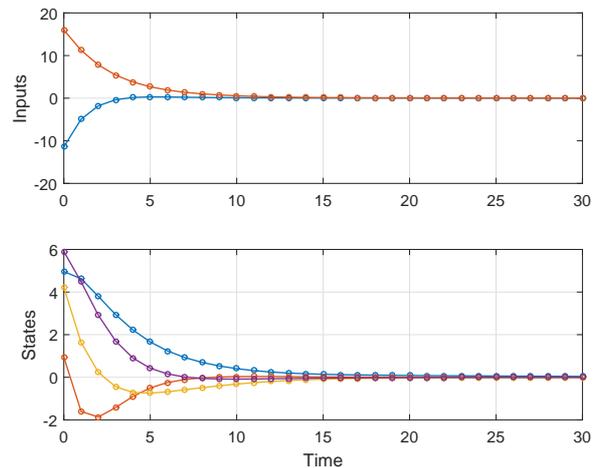}    
		\caption{Optimal input sequence and state response 
		with $N=30$ for the batch reactor system.} 
		\label{fig:states_inputsN30}
	\end{center}
\end{figure}
	
As $N$ grows the solution approximates the infinite-horizon steady-state
solution, which is stabilizing. Figure~\ref{fig:states_inputsN30}
shows the closed-loop response with the data-driven solution
for one experiment carried out with $N=30$.

\section{Conclusions}\label{sec:concl}
We considered a finite-horizon linear quadratic regulation problem, 
where the knowledge about the dynamics of the system is replaced 
by a finite set of input and state data collected from an experiment. 
We have shown that if the experiment is carried out with a sufficiently
exciting input signal then the optimal solution can be computed only 
using the data, with no intermediate identification step,
as the result of a data-dependent semidefinite programming 
problem. 

An important continuation of this research line involves 
the extension of these results to the case where data are affected by noise,
also in comparison with techniques based on system identification \citep{dean2017sample}.
Concerning data-driven methods,
previous efforts in this direction include \cite{cp} 
for stabilization in the presence of input disturbances 
and/or measurement noise, and 
\cite{berberich2019robust}, which considers 
robust performance (including $H_\infty$ control as a special case) 
in the presence of input disturbances.
	
	\bibliographystyle{Harvard}	
	\bibliography{ifacconf}             
	
\end{document}